# Tuning Optoelectronic Properties and Photoelectrochemical Performance of β-TaON via Vanadium Doping

Mirabbos Hojamberdiev[1,2] | Ronald Vargas[3,4] | Lorean Madriz[3,4] | Dilshod Nematov[5,6] | Ulugbek Shaislamov[7] | Hajime Wagata[8] | Yuta Kubota[2] | Kunio Yubuta[9] | Katsuya Teshima[9] | Nobuhiro Matsushita[2]

[1]Mads Clausen Institute, University of Southern Denmark, Sønderborg, Denmark | [2]Department of Materials Science and Engineering, School of Materials and Chemical Technology, Institute of Science Tokyo, Tokyo, Japan | [3]Instituto Tecnológico de Chascomús (INTECH), Consejo Nacional de Investigaciones Científicas y Técnicas (CONICET), Avenida Intendente Marino, Chascomús, Argentina | [4]Escuela de Bio y Nanotecnologías, Universidad Nacional de San Martín (UNSAM), Avenida Intendente Marino, Chascomús, Argentina | [5]Physical–Technical Institute, National Academy of Sciences of Tajikistan, Dushanbe, Tajikistan | [6]School of Optoelectronic Engineering and CQUPT-BUL Innovation Institute, Chongqing University of Posts and Telecommunications, Chongqing, China | [7]Center For Development of Nanotechnology at the National University of Uzbekistan, Tashkent, Uzbekistan | [8]Department of Applied Chemistry, School of Science and Technology, Meiji University, Kawasaki, Japan | [9]Institute for Aqua Regeneration, Shinshu University, Nagano, Japan

**Correspondence:** Mirabbos Hojamberdiev (mirabbos@mci.sdu.dk)

## ABSTRACT

The application of β-TaON for solar-driven water splitting is hindered by limitations in phase purity, stoichiometry, crystallinity, visible-light absorption, carrier mobility, and high recombination rates. This study investigates the impact of vanadium doping (0-25 at.% V) on the structural, optoelectronic, and photoelectrochemical properties of β-TaON using both experimental and density functional theory (DFT) approaches. Phase-pure β-TaON is retained up to 10 at.% V, beyond which secondary phases ($Ta_2O_5$ and VN) form, indicating a threshold of ∼10 at.% under the applied synthesis conditions. All samples exhibit a porous microstructure. Increasing vanadium content induces a redshift in the absorption edge, reducing the bandgap from 2.72 eV (undoped) to 2.38 eV at 25 at.% V for the main β-TaON phase, in agreement with DFT results. X-ray photoelectron spectroscopy confirms substitutional incorporation of $V^{5+}$ for $Ta^{5+}$ in the β-TaON lattice. DFT calculations reveal reduced electron effective mass, enhanced *n*-type conductivity, and favorable band edge shifts enabling spontaneous overall water splitting at ≤10 at.% V. Photoelectrochemical measurements show improved photocurrent and more negative onset potentials for 5–10 at.% V, while higher V doping degrades performance due to phase segregation, which likely increases recombination and hinders interfacial charge transport. Vanadium doping (≤10 at.% V) is an effective strategy for tuning the electronic structure and enhancing the optical properties and photoelectrochemical performance of β-TaON.

## 1 | Introduction

Tantalum oxynitride (TaON) is a representative transition-metal oxynitride known for its advantageous physicochemical characteristics and promising optoelectronic performance [1]. TaON exhibits a diverse range of polymorphic structures: baddeleyite-type (β) [2], $VO_2$(*B*)-type (γ) [3], anatase-type (δ) [4], cotunnite-type [5], bixbyite-type [6], anosovite-type [6], and other theoretically predicted structures [7–10]. Among them, the β-phase has received much attention due to its favorable thermodynamic stability and potential for solar water-splitting applications. Despite reports of a quantum yield reaching 10% for photocatalytic $O_2$ evolution using $\beta$-TaON [11], its practical application in solar water splitting is hindered mainly due to challenges associated with maintaining phase purity and controlling the N:O stoichiometry, both of which critically influence its stability and photocatalytic efficiency.

Cation doping is an effective strategy for modulating the optoelectronic properties and enhancing solar water-splitting activity of transition metal oxynitrides by improving charge separation, light absorption, and surface reaction kinetics [12–14]. In particular, the impact of cation doping on the crystal structure and properties of TaON has also been investigated. For instance, $Mg^{2+}$ doping led to the discovery of a metastable anatase-type phase, exhibiting an orange color, a direct bandgap of 2.2 eV, and anion ordering in three of the *translationengleich* subgroups of $I4_1/amd$ [15]. $Sc^{3+}$ doping resulted in the formation of three distinct structural types (bixbyite, anatase, and anosovite) depending on the Sc content, and metastable anatase-type $Sc_{0.15}Ta_{0.85}O_{1.3}N_{0.7}$ exhibited brilliant colors and a bandgap of 2.54 eV [6]. Guenther and Jansen [16] synthesized $Ta_{(3-x)}Zr_{(x)}N_{(5-x)}O_{(x)}$ ($0 \leq x \leq 0.66$) having the $Ta_3N_5$ structure and $Ta_{(1-x)}Zr_{(x)}N_{(1-x)}O_{(1+x)}$ ($0 \leq x \leq 0.28$) with the TaON structure, and the amounts of Zr and O incorporated into the latter were limited to the small amount. One-step visible-light excitation of overall water splitting over $\beta$-TaON nanoparticles was achieved by Zr doping due to the inhibition of reduced tantalum species, the reduction in particle size, the elimination of grain boundaries, and low defect densities [17]. $Ba^{2+}$ doping did not alter the crystal structure of TaON, and the $Ba^{2+}$-doped TaON 1D array demonstrated a higher photocurrent density than pristine TaON due to the enhanced electrical property [18]. Dronskowski and colleagues [10] provided evidence that $Y^{3+}$-doped TaON may be macroscopically characterized by a cubic fluorite-type defect structure, even though density functional theory (DFT) calculations revealed that each crystallographic unit cell exhibits triclinic distortion. Mimura et al. [19] successfully grew $Y^{3+}$-doped TaON films on (001) YSZ and (012) $Al_2O_3$ substrates, observing that the crystal structure transformed from a monoclinic phase to an orthorhombic phase as the $Y^{3+}$ concentration increased. $La^{3+}$ doping up to 5% in TaON effectively enhanced light absorption and promoted effective separation of photogenerated electrons and holes without forming any secondary phases, improving its photocatalytic activity [20].

Regarding the effect of vanadium doping, a few theoretical studies have been carried out. For instance, Harb and Cavallo [21] theoretically predicted stable monoclinic $Ta_{0.75}V_{0.25}ON$ as a promising photocatalyst for solar water splitting, with a band of 2.0 eV, high absorption efficiency, a static dielectric constant (>10), smaller hole and electron effective masses, and suitable band edge positions. Recently, Jouypazadeh et al. [22] computationally investigated the Nb,V co-doping effect on TaON and found that Nb,V codoped (111) TaON nanosheets are not suitable for overall water splitting because the conduction band edge lies below the $H^+/H_2$ reduction potential; however, they can still be used for the water oxidation half-reaction. Similarly, Lahmer [23] found that S,V-codoped TaON is suitable for the oxygen evolution half-reaction due to its narrow bandgap, enhanced visible light absorption around 700 nm, and reduced work function.

Despite these interesting theoretical findings, the effect of vanadium doping on the optoelectronic and photoelectrochemical performance of $\beta$-TaON has not yet been experimentally studied. In this study, we aim to explore the impact of vanadium doping concentration on phase purity, morphology, optical properties, and charge carrier dynamics of $\beta$-TaON. In addition, density functional theory (DFT) calculations were involved to elucidate the electronic structures, density of states, and carrier effective masses, revealing correlations between structural and electronic properties and photoredox performance. These findings advance the understanding of doping-induced modulation of the optoelectronic properties of $\beta$-TaON, which is critical for enhancing both the efficiency and stability in solar water-splitting systems.

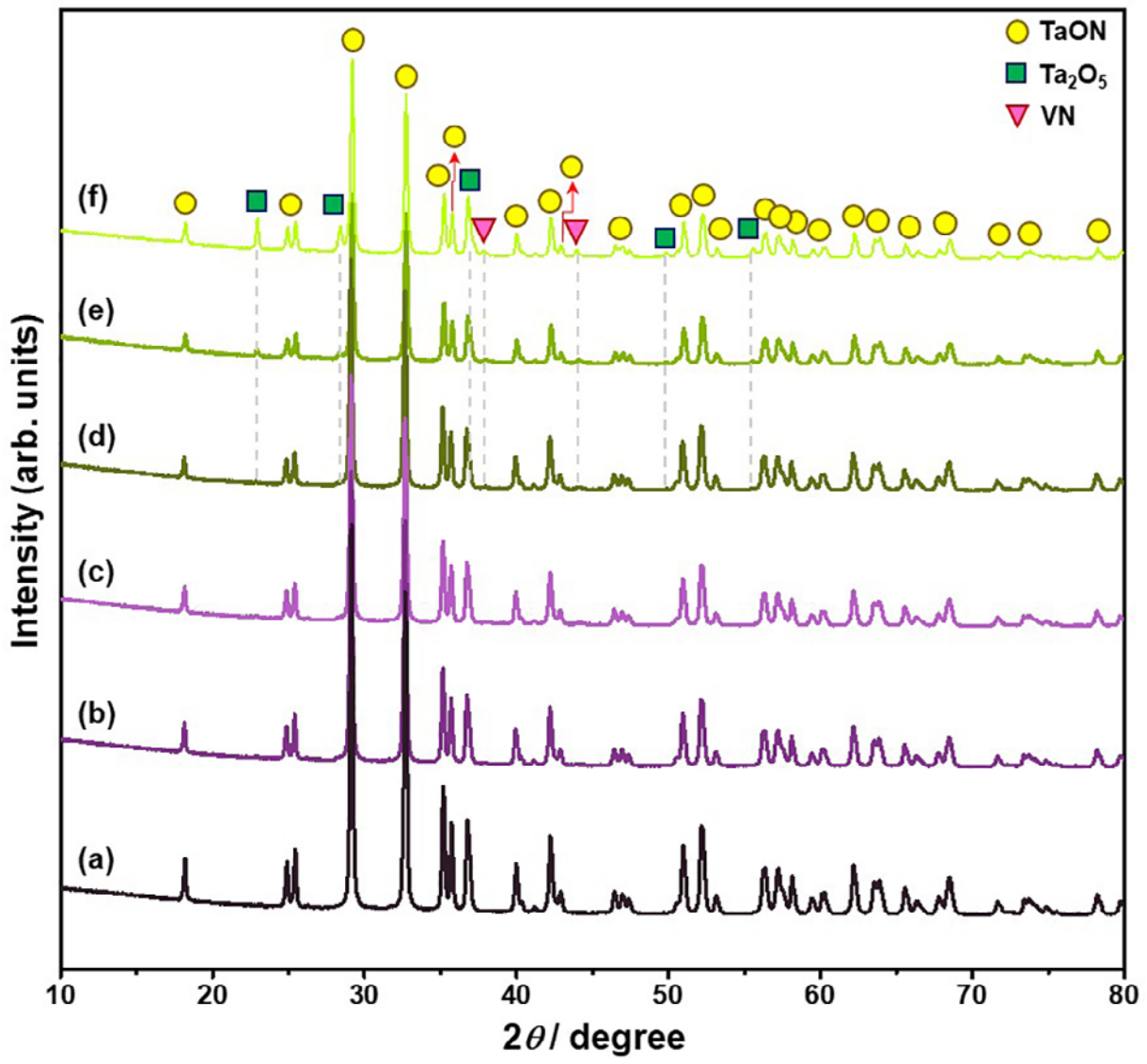

**FIGURE 1** | XRD patterns of pristine and vanadium-doped $\beta$-TaON samples with different vanadium contents: (a) 0 at.%, (b) 5 at.%, (c) 10 at.%, (d) 15 at.%, (e) 20 at.%, and (f) 25 at.%.

## 2 | Results and Discussion

Figure 1 shows the X-ray diffraction (XRD) patterns of pristine and vanadium-doped $\beta$-TaON samples with different vanadium contents (0-25 at.%). The XRD pattern of the undoped sample (TVON0) exhibits well-defined and sharp reflections corresponding to $\beta$-TaON, which crystallizes in a monoclinic structure with space group $P2_1/c$ (no. 14) (ICDD PDF Card No. 98-000-2363), indicative of high crystallinity and phase purity. As the vanadium content increases up to 10 at.%, the $\beta$-TaON phase is preserved with high phase purity, and no secondary phases are formed.

However, a slight shift of reflections toward higher $2\theta$ angles is observed, indicating a reduction in lattice parameters due to the smaller ionic radius of $V^{5+}$ (0.54 Å for CN = 6) compared to $Ta^{5+}$ (0.64 Å for CN = 6). Notably, starting from a vanadium content of 15 at.%, additional reflections emerge, which can be attributed to the secondary phases: $Ta_2O_5$ crystallizing in an orthorhombic structure with space group of $Pmm2$ (No. 25) (ICDD PDF Card No. 98-000-9112) and VN crystallizing in a cubic structure with space group of $Fm\bar{3}m$ (No. 225) (ICDD PDF Card No. 98-064-7603). This suggests that incorporating ≥15 at.% vanadium delays the $Ta_2O_5$-to-$\beta$-TaON conversion under the current ammonolysis conditions and leads to the formation of VN, possibly due to reduced oxygen vacancies, lattice stabilization, increased activation energy for nitridation, decreased diffusion of nitrogen species, and altered surface chemistry, and exceeding the ~10 at.% threshold as a limit under the applied synthesis conditions, which collectively hinder the full nitridation of $Ta_2O_5$ and promote the segregation of VN. Similarly, increasing the amount of tantalum to substitute niobium in $Ba_5Nb_4O_{15}$ partially extended the nitridation duration of the oxide precursors from 20 to 40 h, largely because tantalum forms more stable Ta–O bonds and exhibits a stronger affinity for oxygen than niobium [24].

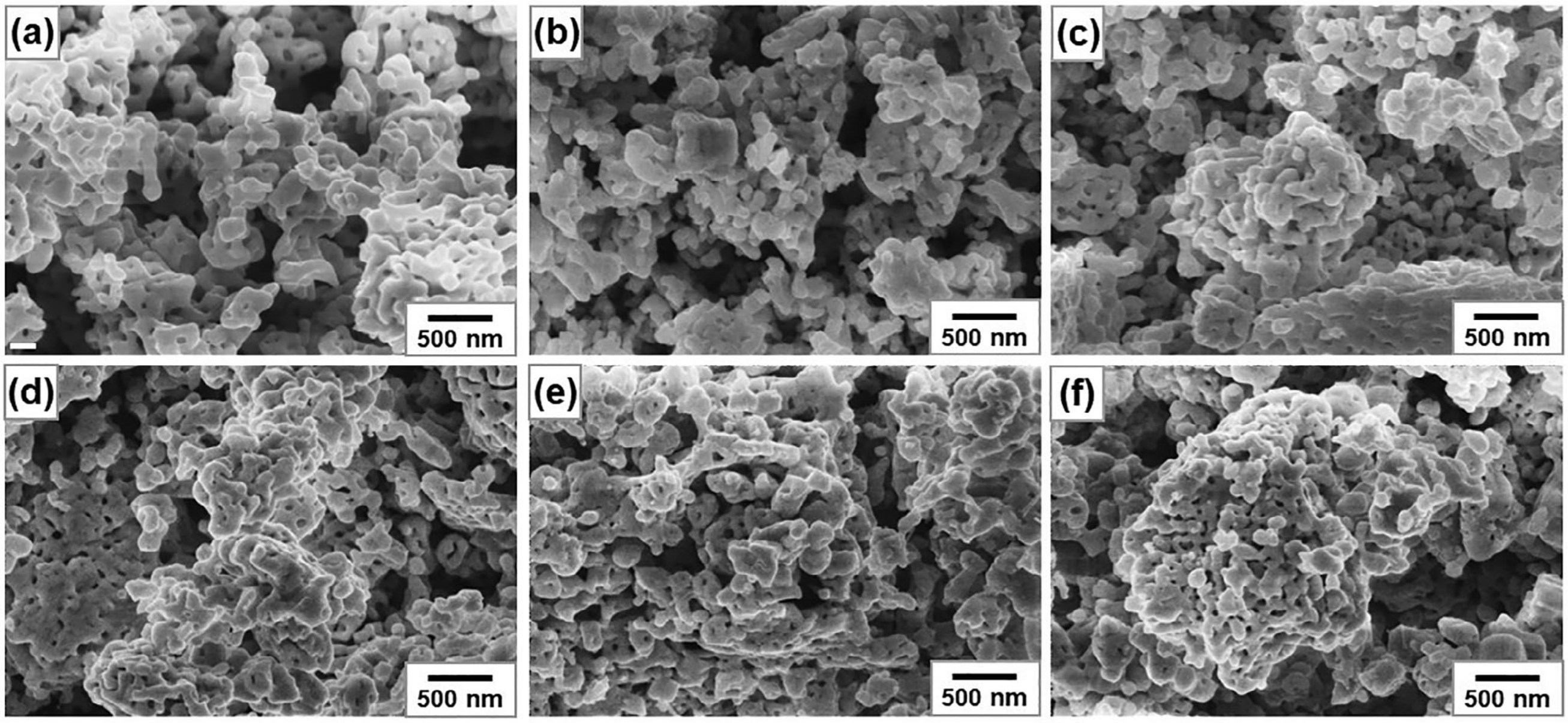

**FIGURE 2** | SEM images of pristine and vanadium-doped $\beta$-TaON samples with different vanadium contents: (a) 0 at.%, (b) 5 at.%, (c) 10 at.%, (d) 15 at.%, (e) 20 at.%, and (f) 25 at.%.

In contrast, Hyuga et al. [25] investigated the effect of various rare-earth oxide dopants on the nitridation behavior of silicon powder, revealing that $Eu_2O_3$ and $CeO_2$ can significantly lower the nitridation onset temperature and enhance its conversion to silicon nitride. Therefore, maintaining vanadium doping below 15 at.% is critical for preserving the single-phase $\beta$-TaON structure under the current ammonolysis conditions. The progressive intensification of the reflections of $Ta_2O_5$ and VN with an increasing vanadium content further confirms the increasing presence of these phases in the synthesized samples. To investigate the effect of vanadium doping on the delay of oxide-to-oxynitride conversion, we also employed different vanadium precursors ($V_2O_4$, VN, and metallic V) under identical ammonolysis conditions. As shown in Figure S1, vanadium doping using $V_2O_4$, VN, and metallic V as the vanadium sources at 25 at.% gradually delayed and completely inhibited the conversion of $Ta_2O_5$ to $\beta$-TaON, respectively. These findings suggest that the ammonolysis conditions must also be adjusted according to both the dopant content and the dopant precursor used.

Figure 2 shows the SEM images of pristine and vanadium-doped $\beta$-TaON samples with different vanadium contents (0-25 at.%). All samples exhibit a porous microstructure consisting of interconnected, irregularly shaped particles. The pristine $\beta$-TaON (Figure 2a) features loosely packed aggregates with minimal surface roughness, whereas lower vanadium contents (5-10 at.%) lead to a more compact structure with slightly roughened surfaces (Figure 2b,c). With the introduction of higher vanadium contents (15-20 at.%), the surface texture evolves into a more irregular form, characterized by denser particle interconnections and the emergence of smaller pores (Figure 2d,e). At a vanadium content of 25 at.%, the sample shows a hierarchical structure composed of densely clustered, rounded nanoparticles, indicating the formation of $Ta_2O_5$ and VN particles along with those of $\beta$-TaON (Figure 2f). The observed morphological change with increasing vanadium content corroborates the XRD results, suggesting a correlation between the formation of secondary phases and the evolution of particle morphology and porosity. Figure S2 shows the EDS elemental mapping images and EDS spectra of pristine and vanadium-doped $\beta$-TaON samples with different vanadium contents (0-25 at.%). In pristine $\beta$-TaON, Ta, O, and N are uniformly distributed, indicating a homogenous phase. In the 5 at.% V-doped $\beta$-TaON, the EDS elemental mapping images confirm the uniform distribution of V along with Ta, O, and N, demonstrating successful incorporation of V into the $\beta$-TaON lattice. As the V content increases from 10 at.% to 25 at.%, the intensity of the V peak in the EDS spectra increases accordingly, reflecting the higher V loading. The pronounced bright regions observed in the EDS elemental mapping images of V and N suggest the segregation of VN at higher V doping content, which is consistent with the XRD results indicating the gradual emergence of VN along with $Ta_2O_5$ secondary phases. The EDS results suggest that V can be effectively doped into the $\beta$-TaON lattice up to a maximum 10 at.% under the current synthesis conditions, beyond which phase segregation becomes significant.

Figure 3 shows the bright-field (BF) TEM and HRTEM images, and SAED patterns of pristine (TVON0) and 25 at.% V-doped (TVON25) $\beta$-TaON samples. The BF-TEM images (Figure 3a,d) of both samples reveal loosely agglomerated nanoparticles with irregular morphologies. In the case of TVON0, the indexed diffraction spots in the SAED pattern (Figure 3b) correspond to the $\beta$-TaON phase, with reflections assigned to the (011), $(\bar{1}11)$, and $(\bar{1}00)$ crystallographic planes. The presence of a twin structure is evident from both the SAED pattern and HRTEM image, suggesting the formation of coherent twin boundaries during crystallization. The HRTEM image of TVON0 (Figure 3c) further shows well-defined and continuous lattice fringes extending across the entire nanoparticle, indicative of a high degree of crystallinity and the absence of dislocations or significant structural defects. The diffraction spots in the SAED pattern of TVON25 (Figure 3e) can be indexed to the reflections from the $(0\bar{1}2)$, $(1\bar{1}2)$, and $(\bar{1}00)$ crystallographic planes of the $\beta$-TaON phase, confirming the retention of the crystal structure upon doping. The observed lattice spacing of 0.486 nm corresponds to the (100) crystallographic plane of the $\beta$-TaON phase. Similarly, the HRTEM image of TVON25 (Figure 3f) shows clearly resolved lattice spacing corresponding to the (100) crystallographic plane of the $\beta$-TaON phase. Figure 4 shows the HAADF-STEM image together with the corresponding STEM-EDS elemental distribution maps of the TVON25 sample. The elemental maps clearly reveal the spatial distribution of tantalum, oxygen, nitrogen, and vanadium, confirming $\beta$-TaON as the primary phase and identifying $Ta_2O_5$ and VN as secondary crystalline phases.

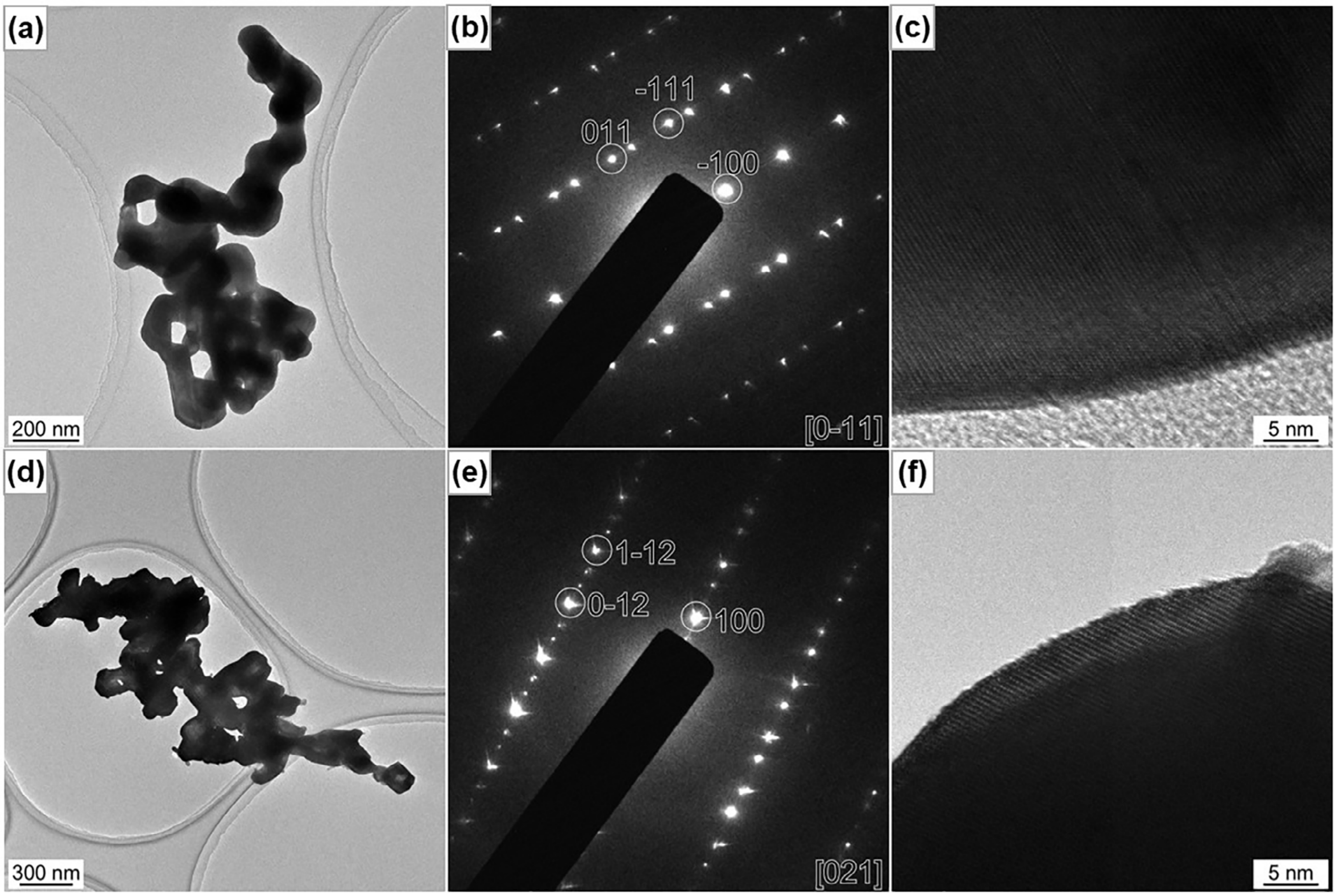

**FIGURE 3** | Bright-field TEM images, SAED patterns, and HRTEM images of pristine (a-c) and 25 at.% V-doped (d-f) $\beta$-TaON samples.

Figure 5 shows the high-resolution XPS spectra of the Ta 4f, N 1s, O 1s, and V 2p core levels for the TVON0 and TVON10 samples. The Ta 4f spectra were deconvoluted using Gaussian fitting, revealing two characteristic peaks at 28.1 and 26.2 eV, corresponding to the Ta $4f_{5/2}$ and Ta $4f_{7/2}$ spin-orbit components, respectively. The spin-orbit splitting of 1.9 eV is consistent with the +5 oxidation state of tantalum and agrees well with previously reported values (28.2 and 26.3 eV, $\Delta E \approx 1.9$ eV) [26]. Across the series, no significant shifts in the binding energies of the Ta 4f peaks were detected, suggesting that the oxidation state of tantalum remains unaltered upon vanadium doping. However, a minor decrease in Ta 4f peak intensity in the 10 at.% V-doped sample (TVON10) indicates a partial substitution of $Ta^{5+}$ by $V^{5+}$, which is in good agreement with earlier reports [13]. The N 1s spectra exhibit an intense peak centered at 397.4 eV, which is characteristic of the Ta–N bond [27]. Deconvolution reveals a second, minor component at 396.2 eV, which could be attributed to N atoms adjacent to oxygen vacancies. Upon vanadium doping, both peaks shift to lower binding energy (397.2 and 395.7 eV), and the intensity of the defect-related component is substantially reduced. This coordinated shift indicates an overall increase in electron density on nitrogen sites, while the suppression of the minor peak suggests vanadium ions actively passivate the associated vacancy complexes. Deconvolution of the O 1s spectra reveals a main peak at ~530.8 eV, corresponding to lattice oxygen (Ta–O), and a secondary component in the 532.4–533 eV range. This higher binding energy component is consistent with the presence of surface species, such as hydroxyl groups (—OH), and may also include contributions from adsorbed water or carbon-oxygen contaminants. As expected, the V 2p region shows no detectable signal for TVON0. For the 10 at.% V-doped sample (TVON10), the V 2p core-level spectrum was fitted using two chemically distinct spin-orbit doublets. The V $2p_{3/2}$ components are located at 517.3 and 515.6 eV, which are assigned to $V^{5+}$ and $V^{4+}$ oxidation states, respectively. The corresponding V $2p_{1/2}$ components appear at 524.7 and 523.2 eV, giving a characteristic spin-orbit splitting of approximately 7.4 eV with the intensity ratio constrained to 2:1, as expected for vanadium 2p levels [28, 29]. In addition, a broad feature centered at ∼521-522 eV was included in the fitting and attributed to the $O_{KLL}$ X-ray satellite of the O 1s core level [30–32]. The dominance of the $V^{5+}$ $2p_{3/2}$ component indicates that vanadium is present predominantly in the +5 oxidation state, supporting its isovalent substitution for $Ta^{5+}$ in the β-TaON lattice.

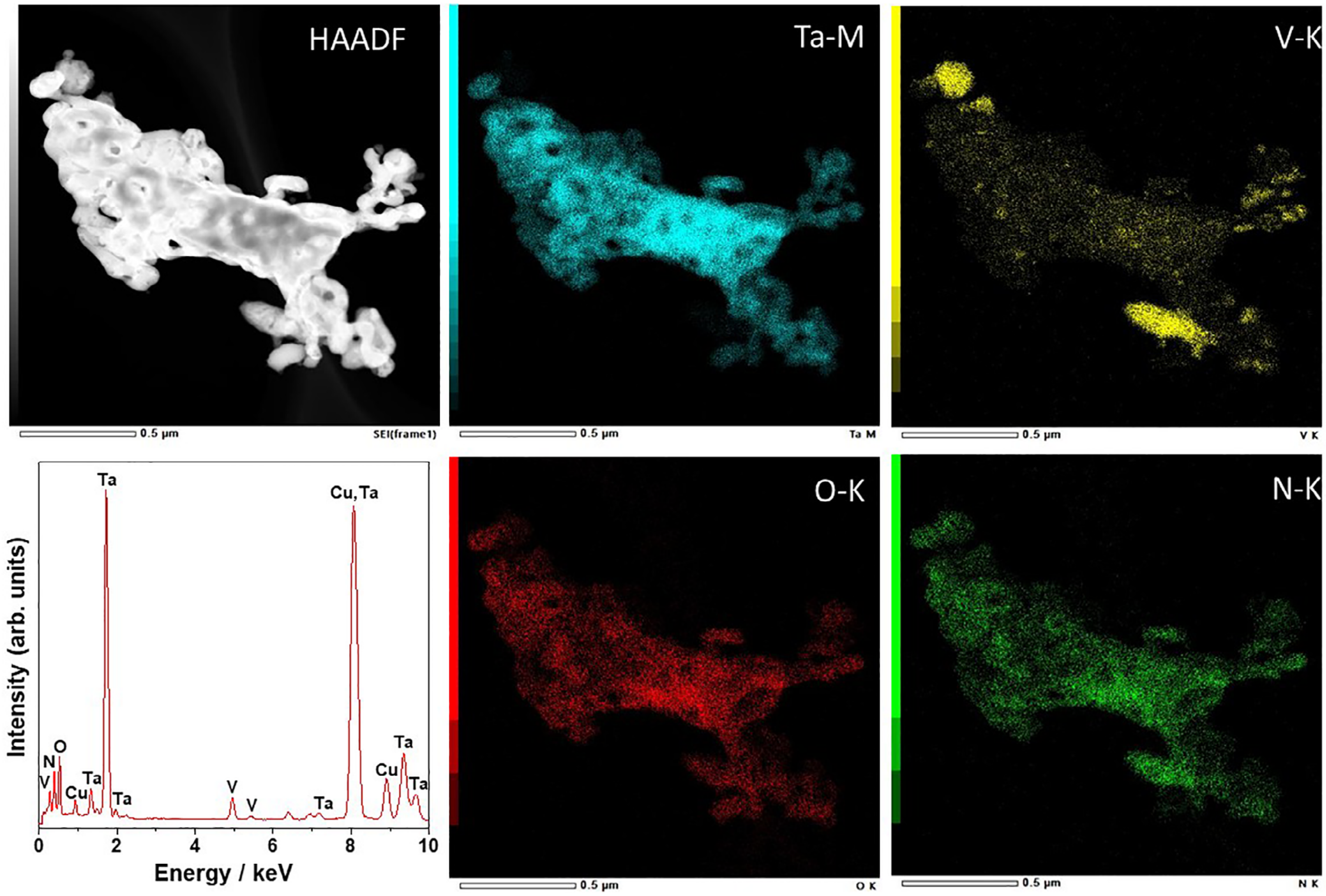

**FIGURE 4** | HAADF-STEM image and corresponding STEM-EDS elemental distribution maps of TVON25.

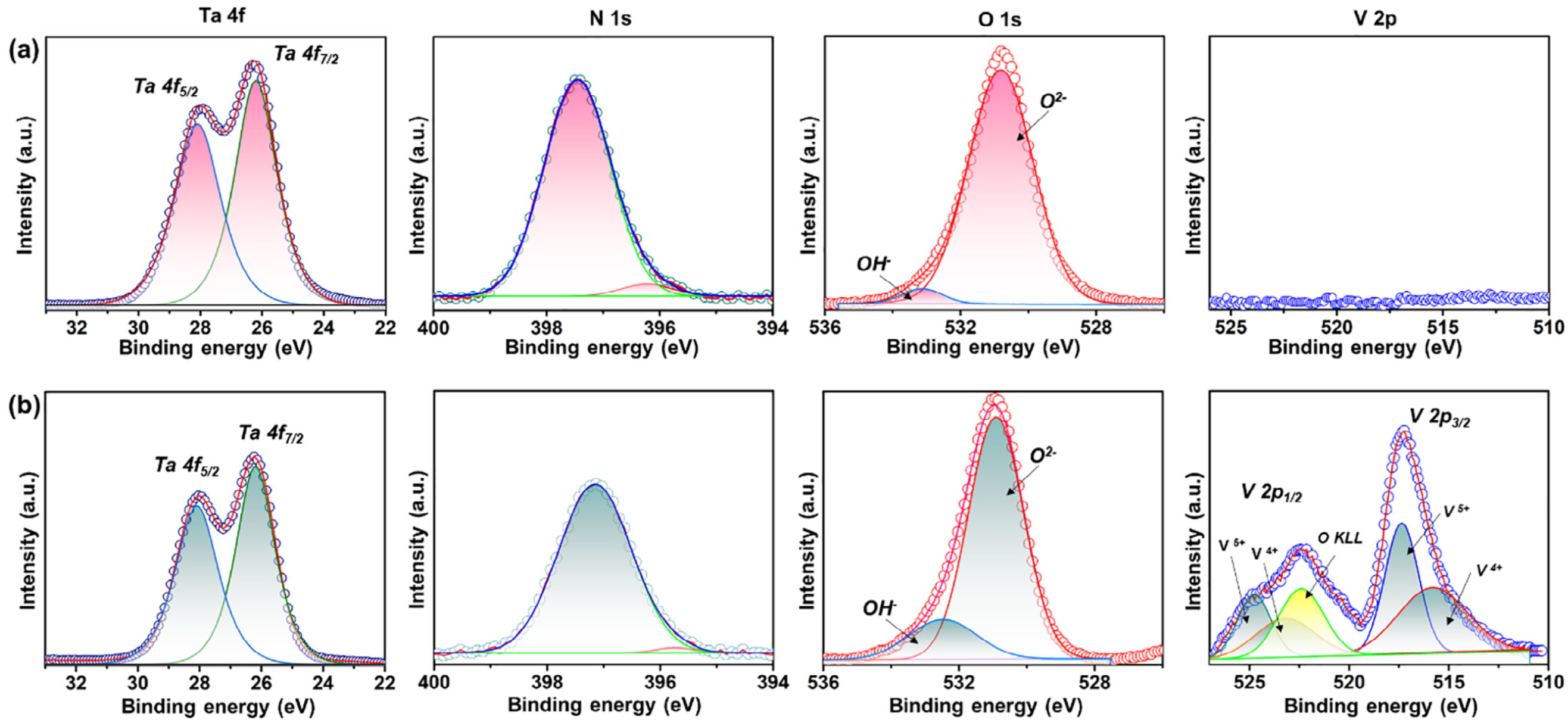

**FIGURE 5** | High-resolution XPS spectra of the Ta 4f, N 1s, O 1s, and V 2p for pristine (a) and 10 at.% V-doped (b) $\beta$-TaON samples. Peak deconvolutions and assignments are indicated.

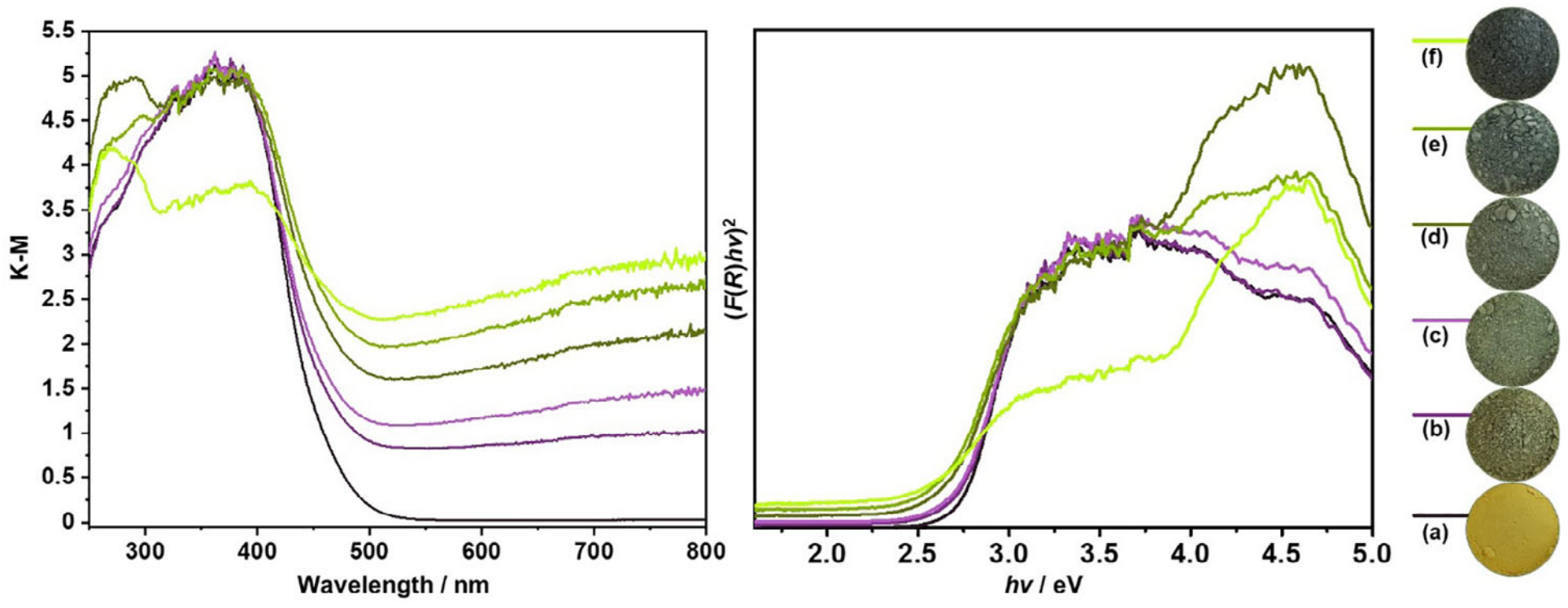

**FIGURE 6** | UV–vis diffuse reflectance spectra (*left*) and corresponding Tauc plots (*right*) of pristine and vanadium-doped β-TaON samples with different vanadium contents: (a) 0 at.%, (b) 5 at.%, (c) 10 at.%, (d) 15 at.%, (e) 20 at.%, and (f) 25 at.%.

Figure 6 presents the UV–vis diffuse reflectance spectra and corresponding Tauc plots of pristine and vanadium-doped β-TaON samples with different vanadium contents (0-25 at.%). The UV–Vis diffuse reflectance spectra reveal a pronounced redshift of the absorption edge as the vanadium content increases, indicating a progressive narrowing of the optical bandgap. Alongside this redshift, a notable enhancement in absorption beyond the absorption edge is observed, particularly at higher vanadium contents. This enhanced absorption can be attributed to the introduction of localized defect states within the bandgap, likely stemming from charge-compensating defects associated with vanadium substitution [33, 34]. Also, for the samples with higher vanadium content (>10 at.%), an additional absorption shoulder is observed. This feature appears at shorter wavelengths, which is in good agreement with the expected wide bandgap of the secondary $Ta_2O_5$ phase identified in the XRD patterns. The Tauc plots, constructed under the assumption of a direct allowed electronic transition, show a gradual decrease in the optical bandgap from approximately 2.72 eV (pristine) to 2.38 eV with increasing vanadium content for the main β-TaON phase. This bandgap narrowing suggests substantial modification of the electronic structure of β-TaON, plausibly due to the hybridization of V 3d orbitals with the Ta 5d–O 2p/N 2p states, as well as the formation of defect levels. For the samples containing 15 at.%, 20 at.%, and 25 at.% V, the observed secondary absorption edges were estimated to be 3.27, 3, and 3.5 eV, respectively, which are close to the reported bandgap value of $Ta_2O_5$. Corroborating the spectroscopic data, a visible change in powder color is also observed with increasing vanadium content, from pale yellow (pristine) to light grey, and ultimately to dark grey, reflecting the enhanced visible-light absorption and reduced bandgap.

The structural optimization of pristine and vanadium-doped β-TaON with different vanadium contents (0-25 at.%), performed using the SCAN functional, shows a gradual decrease in the unit cell volume with increasing vanadium content (Figure 7a). This typical behavior is attributed to the partial substitution of the larger $Ta^{5+}$ ion (0.64 Å for CN = 6) with the smaller $V^{5+}$ ions (0.54 Å for CN = 6), in agreement with Vegard's law [35]. The absence of any discontinuities or anomalies in the unit cell volume across the compositional range indicates uniform incorporation of vanadium into the β-TaON lattice, revealing its structural stability. The electronic properties were investigated using the Tran-Blaha modified Becke–Johnson (TB-mBJ) potential. Figure 7b shows the variation in the bandgap as a function of vanadium content. A systematic narrowing of the calculated bandgap (TB-mBJ, theoretical values) is observed, decreasing from 3.10 eV for TVON0 to 1.68 eV for TVON25. For pristine β-TaON (TVON0), the SCAN→TB-mBJ scheme yields a direct bandgap of 3.10 eV, which is in good agreement with previous HSE06 hybrid calculations reporting 3.0-3.02 eV for β-TaON [21, 36]. These theoretical values differ from the experimental optical bandgap values obtained from Tauc diagrams for the main β-TaON phase (2.72–2.38 eV) due to methodological differences between optical absorption measurements and electronic structure calculations. This reduction is attributed to the incorporation of V 3d states into the conduction band, which effectively lowers the conduction band minimum. The values of the calculated bandgap, charge carrier effective masses, and Fermi energy are summarized in Table 1. With increasing vanadium content, the electron effective mass decreases from 0.49 to 0.36 $m_e$, while the hole effective mass shows a slight increase. The upward shift of the Fermi level may indicate enhanced density of states near the conduction band edge and an increase in *n*-type conductivity. This trend in the electronic structure is consistent with earlier studies on vanadium-doped $SrTiO_3$ and TaON [21, 37], where vanadium doping led to a reduced bandgap and improved optical absorption.

Further insights into the impact of vanadium doping on the electronic structure are provided by the total density of states (TDOS) analysis. As the vanadium content increases, a pronounced narrowing of the bandgap is observed, agreeing well with previous findings (Figure 7c). A significant increase in the density of states near the Fermi level is attributed to the contribution of V 3d orbitals, facilitating enhanced electronic transitions within the visible range. Importantly, all compositions retain semiconducting behavior, with the Fermi level located within or near the band edges, and no evidence of a metallic phase transition is observed. The partial density of states (PDOS) diagrams, shown in Figure S3, reveal that the valence band is predominantly composed of O 2p and N 2p states, while the conduction band becomes progressively enriched with vanadium-derived states as the doping level increases. This enhances optical transitions and contributes to increased electron mobility. A similar effect was reported for V-doped $SrTiO_3$ [37], although in the present case, the semiconducting nature is preserved across all compositions.

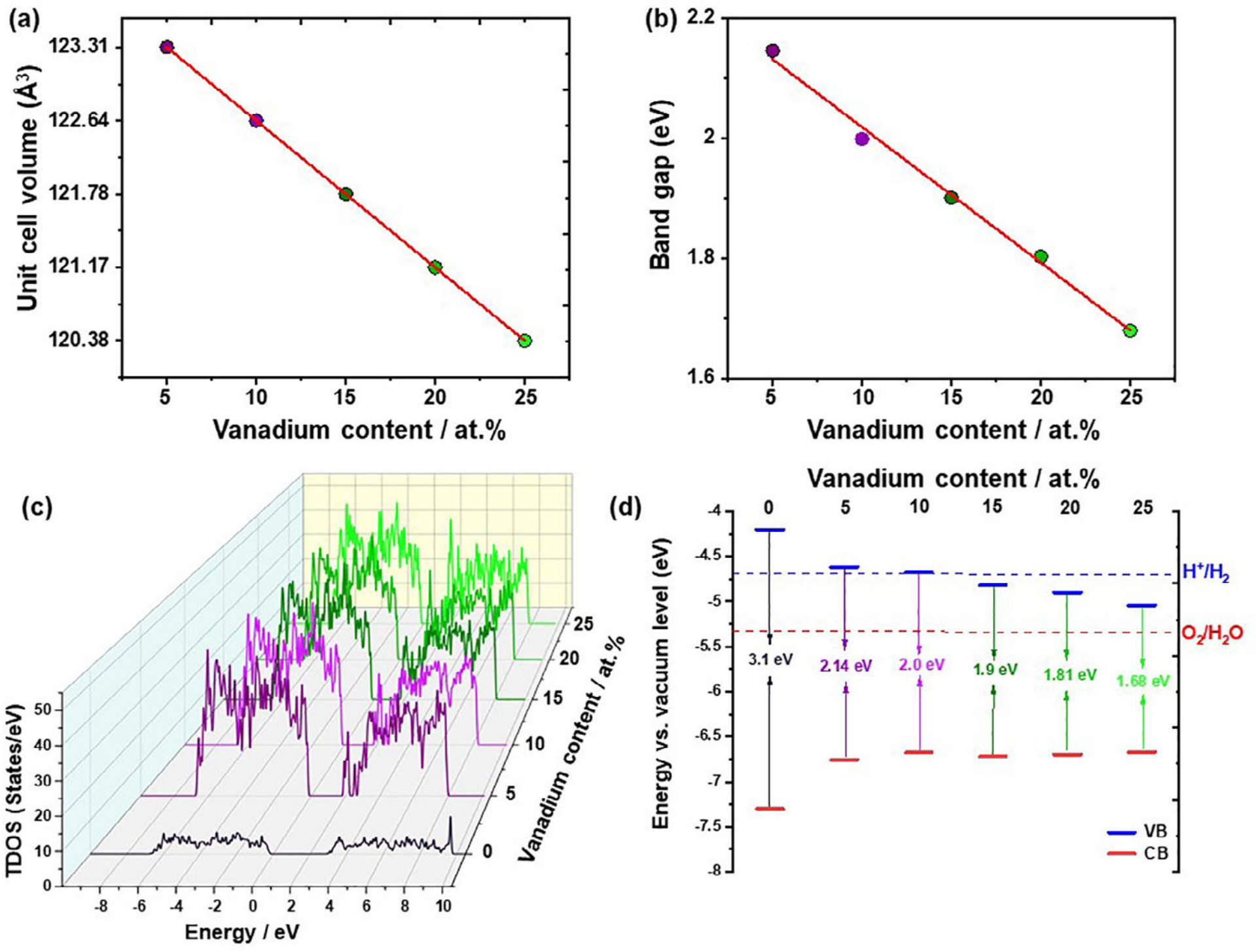

**FIGURE 7** | Unit cell volumes (a), bandgaps (b), total density of states (c), and absolute positions of the valence band maximum and conduction band minimum (d) of pristine and vanadium-doped $\beta$-TaON samples with different vanadium contents: 0 at.%, 5 at.%, 10 at.%, 15 at.%, 20 at.%, and 25 at.%.

**TABLE 1** | Fermi energy (EF), bandgap ($E_g$), ratios of effective masses of holes ($m_h*/m_0$) and electrons ($m_e*/m_0$) relative to the free electron mass $m_0$, refractive index ($n$) in the low-frequency limit, and static permittivity ($\varepsilon$) of pristine and vanadium-doped $\beta$-TaON with different vanadium contents (0–25 at.%).

| Vanadium content, at. % | $E_F$, eV | $E_g$, eV | $m_h^*/m_0$ | $m_e^*/m_0$ | $n$ | $\varepsilon$ |
|---|---|---|---|---|---|---|
| 0 | 6.88 | 3.10 | 0.42 | 0.31 | 2.71 | 7.46 |
| 5 | 6.82 | 2.14 | 0.47 | 0.36 | 3.26 | 10.64 |
| 10 | 6.81 | 2.00 | 0.50 | 0.40 | 3.32 | 11.06 |
| 15 | 6.90 | 1.90 | 0.53 | 0.44 | 3.44 | 11.85 |
| 20 | 6.93 | 1.80 | 0.55 | 0.46 | 3.51 | 12.33 |
| 25 | 6.99 | 1.68 | 0.60 | 0.50 | 3.63 | 13.22 |

Figure 7d shows the absolute positions of the valence band maximum (VBM) and conduction band minimum (CBM) for compositions relative to the vacuum level. For reference, the standard redox potentials for hydrogen evolution ($H^+/H_2$, 0 V vs. NHE) and oxygen evolution ($O_2/H_2O$, 1.23 V vs. NHE) are also indicated. For TVON0, TVON5, and TVON10, the valence band (VB) lies at a more positive potential than the standard water oxidation potential ($O_2/H_2O$, 1.23 V vs. RHE), while the conduction band (CB) remains more negative than the hydrogen evolution reaction (HER, 0 V vs. RHE). This configuration ensures that both half-reactions are thermodynamically allowed, enabling full water splitting under suitable kinetic conditions. However, as the vanadium content increases (TVON15 and TVON25), the conduction band shifts to more positive potentials, eventually crossing the HER threshold. Although the valence band remains sufficiently positive to oxidize water, the less negative CB edge compromises the ability to reduce protons. Therefore, for TVON15, TVON20, and TVON25, overall water splitting is thermodynamically unfeasible, and only oxidative half-reactions (e.g., organic pollutant degradation or $O_2$ evolu-tion) are supported. These findings are in good agreement with previous studies on related oxynitrides, such as $YTaON_2$ and $Ta_3N_5$ [22, 38].

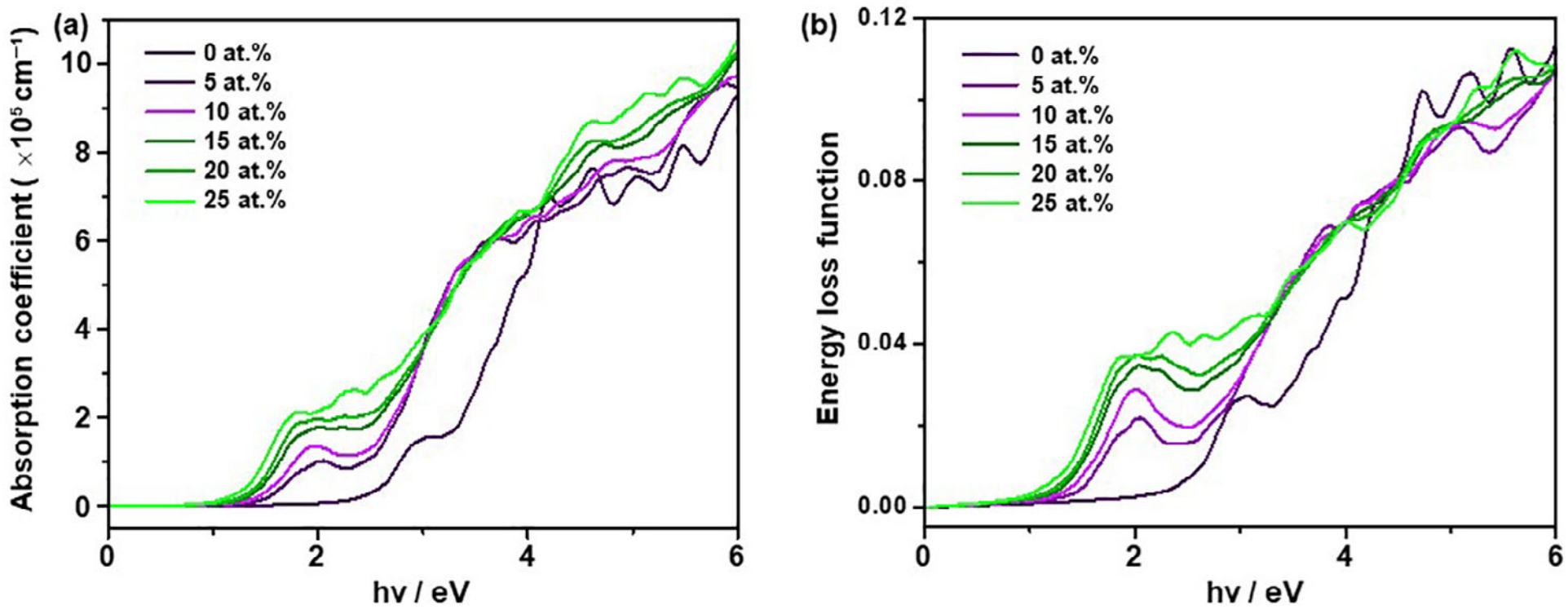

**FIGURE 8** | Absorption coefficient (a) and energy loss function (b) of pristine and vanadium-doped $\beta$-TaON samples with different vanadium contents: 0 at.%, 5 at.%, 10 at.%, 15 at.%, 20 at.%, and 25 at.%.

The optical properties were evaluated using the real ($\varepsilon_1$) and imaginary ($\varepsilon_2$) components of the dielectric function. As shown in Figure 8a, the absorption coefficients exhibit a pronounced redshift in the absorption edge with increasing vanadium content, indicating an extension of the absorption spectrum into the visible range. This trend is consistent with the observed bandgap narrowing shown in Table 1. The most significant enhancement in visible-light absorption is observed for compositions with 20 at.% and 25 at.% V, where the absorption range extends up to approximately 600–800 nm. These results support the potential of V-doped $\beta$-TaON as an efficient photocatalyst under visible light irradiation.

Figure 8b shows the energy loss function, which represents the collective electronic excitations induced by fast-charged particles. The primary plasmon peak exhibits a gradual shift toward lower energies with increasing vanadium content, indicating a reduction in plasmon energy and an enhancement in polarizability. This trend may also be indicative of increased electrical conductivity and higher free carrier concentration. As summarized in Table 1, the calculated optical constants reveal that the real part of the dielectric permittivity ($\varepsilon_1$) increases in the low-frequency region with increasing vanadium content, accompanied by a corresponding increase in the refractive index ($n$). These observations suggest an enhanced dielectric response, improved electrostatic screening, and potentially more efficient separation of photogenerated charge carriers.

The phonon density of states shown in Figure S4a exhibits no negative (imaginary) frequencies, indicating the absence of pronounced dynamical instabilities at 0 K. This observation is consistent with previously reported DFT studies on related oxynitrides and nitrides [39]. Furthermore, the phonon dispersion curves calculated for $\beta$-TaON (TVON0) and 10 at.% V-doped TaON (TVON10), presented in Figure S5a,b, also shows no imaginary frequencies, providing additional confirmation that no strong dynamical instabilities are present in the examined structures under the applied computational conditions. The temperature-dependent thermodynamic functions of the compositions, including entropy ($S$), Helmholtz free energy ($F$), and constant-volume heat capacity ($C_v$), are presented in Figure S4b–d. All trends are physically consistent: entropy and heat capacity increase with temperature, while free energy decreases. This behavior supports the thermodynamic stability of the system, particularly near room temperature, where entropy becomes a stabilizing factor in accordance with thermodynamic principles. Additionally, the heat capacity values approach the Dulong-Petit limit at temperatures above 700 K, further confirming the reliability of the calculations.

Following structural and optoelectronic analyses, photoelectrochemical (PEC) studies were carried out to assess the effect of vanadium incorporation into $\beta$-TaON. The TVON0, TVON5, TVON10, TVON15, TVON20, and TVON25 samples exhibited measurable PEC response, although with key differences arising from their distinct semiconductor-electrolyte interfacial behavior. To evaluate the PEC activity of these photocatalysts, open-circuit potential (OCP) measurements were performed at different levels of irradiance after reaching steady state. This method serves as a sensitive probe for the photoinduced charge separation capability of photoanodes [40]. As shown in Figure 9a, all synthesized samples exhibited a shift in OCP toward more negative values with increasing light intensity, indicating the accumulation of photogenerated electrons in the semiconductor. The OCP response to irradiance provides key insights into the charge separation efficiency and Fermi level dynamics at the semiconductor surface under illumination. For pristine $\beta$-TaON (TVON0), the OCP shift with light intensity is moderate, suggesting a relatively balanced rate of photogenerated charge accumulation and recombination. Upon substitutional incorporation of vanadium (TVON5 and TVON10), a more pronounced negative OCP shift is observed at higher irradiance, implying that local chemical environments influence interfacial energetics and contribute to the enhanced photoresponse. This behavior can be attributed to enhanced upward band bending at the semiconductor-electrolyte interface, indicating more efficient separation of photogenerated charge carriers and stabilization of the *quasi*-Fermi level for electrons. This interpretation is consistent with the XPS results, which show shifts in the Ta 4f and O 1s binding energies, and with DFT calculations suggesting modulation of the electronic density near the conduction band minimum. Similar to V-doped $SrTiO_3$, dopant states overlap the conduction band, reducing recombination centers and enhancing carrier transport [37]. For TVON15, TVON20, and TVON25, the OCP trend becomes less favorable, showing reduced shifts or even plateaus at higher irradiance. This suggests saturation of surface states or increased recombination due to the formation of additional dispersed phases, as supported by TEM/EDS and XRD results indicating crystallographic disorder and secondary phase segregation ($Ta_2O_5$ and/or VN). Vanadium vacancy-related defects are known to introduce trap states that promote recombination, as observed in related oxides [41].

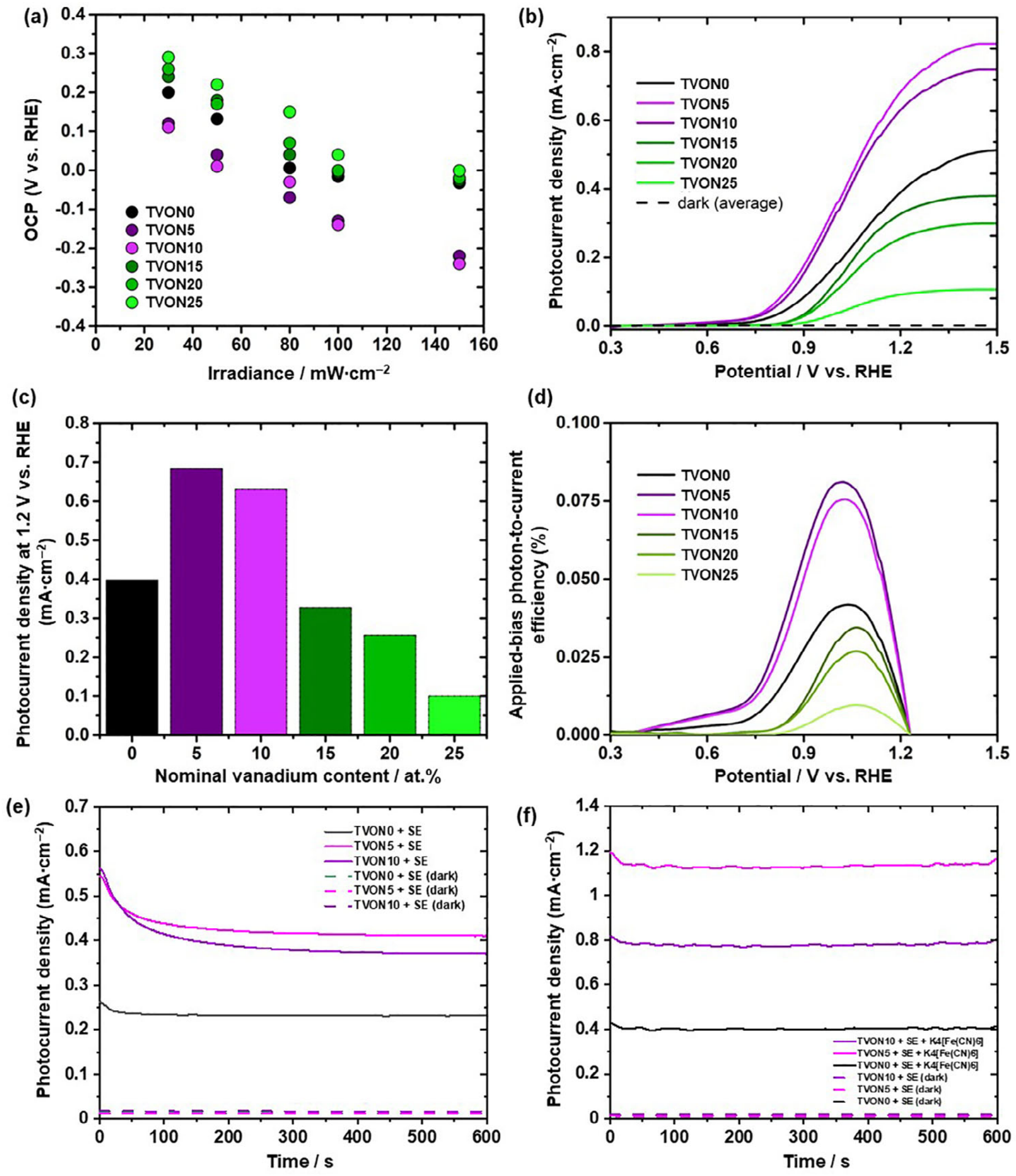

**FIGURE 9** | Plot of OCP vs. irradiance (relative error ~3% from duplicates) (a), LSV (b), current density (at 1.2 V vs. RHE) at different nominal vanadium doping concentrations (relative error ~4% from duplicates) (c), and ABPE vs. potential (d) of TVON0, TVON5, TVON10, TVON15, TVON20, and TVON25. CA tests at 1 V vs. RHE: in supporting electrolyte (SE) (e) and in SE + potassium ferrocyanide ($K_4[Fe(CN)_6]$) (f) of TVON0, TVON5, and TVON10. Data in (b–f) were measured at 100 mW•cm$^{-2}$.

Linear sweep voltammetry (LSV) under illumination (Figure 9b) shows the PEC response of all samples, with the sustained photocurrent at the interface primarily attributed to the water oxidation reaction (WOR) [42]. The LSV profiles under illumination show a clear correlation between V content and photocurrent density, as also reflected in the trend of the current density at 1.2 V *vs*. RHE at different nominal vanadium doping concentrations (Figure 9c). Pristine $\beta$-TaON shows a photocurrent equal to 0.389 mA cm$^{-2}$ at 1.2 V *vs*. RHE and a high onset potential (0.8 V *vs*. RHE), indicating relatively slow hole-transfer kinetics, which is consistent with our previous result [24]. Upon introduction of 5 at.% and 10 at.% vanadium (TVON5 and TVON10), the onset potential decreases to ~0.75 V *vs*. RHE for both samples, and the photocurrent increases notably, 0.683 mA cm$^{-2}$ for TVON5 and 0.63 mA cm$^{-2}$ for TVON10. This improvement suggests enhanced hole transport and potentially faster WOR kinetics, driven by reduced recombination, which is consistent with the electronic activation of surface Ta-O-V sites observed by XPS, as well as DFT results indicating bandgap narrowing and localized acceptor states near the valence band. These effects resemble enhancements reported for V-doped $Fe_2O_3$, where band-gap narrowing and increased donor density lead to higher photocurrents [43].

Reported photocurrents for TaON have varied widely in the literature due to differences in stoichiometry control, film architecture, and deposition methodology. For powder-based TaON photoanodes prepared by conventional ammonolysis, typical photocurrent densities lie within 0.1-1.0 mA•cm$^{-2}$ at ~1.0-1.23 V *vs*. RHE. In contrast, highly optimized TaON films with precisely controlled O/N ratios have achieved ~1.0-2.0 mA•cm$^{-2}$ at 1 V *vs*. RHE [24, 44, 45]. Under comparable powder-based conditions, the photocurrents obtained in this work (0.389 mA•cm$^{-2}$ for pristine $\beta$-TaON and 0.63–0.68 mA•cm$^{-2}$ for V-doped samples) fell within the expected range for non-engineered TaON electrodes. The enhancement induced by moderate vanadium incorporation was therefore significant and consistent, while remaining below the performance of optimized and architectured TaON films.

In contrast, TVON15, TVON20, and TVON25 show a marked decrease in photocurrent (0.327 mA•cm$^{-2}$ for TVON15, 0.256 mA•cm$^{-2}$ for TVON20, and 0.091 mA•cm$^{-2}$ for TVON25 at 1.2 V *vs*. RHE), with onset potential (~0.9 V *vs*. RHE for TVON15 and TVON20 and ~0.98 V vs. RHE for TVON25) more positive than that of undoped $\beta$-TaON. These effects are attributed to increased bulk and interfacial recombination, likely caused by deep-level trap states introduced by excess vanadium. The broadened valence-band features in XPS and the electronic disorder revealed by the calculated PDOS support this interpretation. Furthermore, the possible formation of a poorly integrated, vanadium-rich secondary phase may further hinder charge transport and suppress interfacial charge transfer. Similar trends have been reported for $WO_3$ doped with excess vanadium, where optimal doping enhances activity, while excessive doping introduces recombination centers and lattice defects [46].

The applied-bias photon-to-current efficiency (ABPE) curves (Figure 9d) exhibit a downward concave shape with maxima between 1.0 and 1.1 V *vs*. RHE. This behavior aligns with trends reported for undoped $\beta$-TaON synthesized under optimized ammonolysis conditions, where precise control of oxygen partial pressure during ammonolysis plays a critical role in tuning its photoelectrochemical response [24]. TVON0, TVON5, and TVON10 exhibit distinct maxima in their ABPE *versus* potential curves, with TVON0 reaching a maximum of 0.042%, TVON5 achieving 0.081%, and TVON10 reaching 0.075%. Notably, the peak of TVON5 and TVON10 significantly exceeds that of TVON0 and all other samples (0.034% for TVON15, 0.026% for TVON20, and 0.010% for TVON25), highlighting their superior PEC activity among the photocatalysts studied. This reinforces the notion that moderate vanadium doping optimally balances carrier generation and transport while minimizing recombination.

To further probe carrier dynamics in the most representative electrodes, chronoamperometry (CA) experiments were incorporated as a qualitative diagnostic. CA under continuous illumination was performed on the TVON0, TVON5, and TVON10 samples (Figure 9e,f), which were selected because they define the most relevant photocurrent values within the vanadium doping series. Upon illumination in the supporting electrolyte, all three electrodes exhibited an initial high photocurrent transient that decayed rapidly to a *quasi*-steady-state value, a behavior characteristic of surface charge accumulation and surface recombination processes in semiconductor photoanodes [47]. Upon addition of potassium ferrocyanide, the transient overshoot was markedly suppressed, and the steady-state photocurrent increased, reflecting more efficient hole extraction and reduced interfacial recombination due to the presence of a fast outer-sphere electron-transfer chemical specie [24, 47, 48]. These qualitative signatures are consistent with the recombination-limited kinetics inferred from the LSV analysis and with the enhanced interfacial charge-transfer behavior observed for the moderately V-doped samples, supporting the expected balance between interfacial charge transfer and recombination in semiconductor electrodes. In addition, the steady-state current remains stable over the explored time window, and in both the supporting electrolyte and in the presence of potassium ferrocyanide, the photocurrents are consistently higher than under dark conditions.

The enhanced PEC performance of TVON5 and TVON10 can be attributed to the synergy between their optimized average structural and electronic band structures. XRD analysis indicates that at low V concentrations (5-10 at.%), no secondary crystalline phases are detected and that the $\beta$-TaON structure is preserved. These observations are consistent with vanadium incorporation into the host lattice as inferred from the absence of detectable secondary crystalline phases; however, the formation of a solid solution cannot be conclusively confirmed, and nanoscopic compositional inhomogeneities also cannot be excluded. Guided by this structural picture, DFT calculations suggest that vanadium incorporation introduces local electronic perturbations, leading to band-gap narrowing through V 3d–N 2p hybridization. This modification enhances visible-light absorption while maintaining a favorable valence-band edge position, which is critical for driving the water oxidation reaction (WOR). Simultaneously, the conduction band remains sufficiently negative to satisfy the thermodynamic requirement for charge separation. The net effect is a reduced recombination probability and enhanced separation of photogenerated charge carriers across the semiconductor-electrolyte interface. Experimentally, this manifests as a higher photocurrent density and an earlier onset potential under illumination, with TVON5 demonstrating the optimal balance between band alignment, structural coherence, and optoelectronic activation, followed closely by TVON10. This balance is delicate: excessive vanadium doping beyond 10 at.% leads to lattice destabilization and defect-induced recombination, ultimately degrading PEC efficiency. At higher vanadium concentrations (TVON15, TVON20, and TVON25), PEC performance declines due to the formation of secondary phases and increased structural disorder. The emergence of $Ta_2O_5$ and VN phases disrupts lattice uniformity and introduces recombination centers. Particularly, $Ta_2O_5$ is primarily UV-active, limiting visible light utilization. DFT results confirm the presence of mid-gap states and deep traps that reduce charge carrier lifetimes and hinder effective water oxidation, offsetting any benefits gained from improved light absorption.

## 3 | Conclusion

In summary, the effect of vanadium doping content on structural, optical, and photoelectrochemical properties of $\beta$-TaON was studied. The structural analysis showed that phase-pure $\beta$-TaON could be maintained up to 10 at.% vanadium doping, indicating the ~10 at.% threshold as a limit under the applied synthesis conditions. Vanadium doping resulted in the narrowing of the optical bandgap from 2.72 to 2.38 eV for the main $\beta$-TaON phase and an extended absorption edge into the visible-light region, which was also complemented by DFT calculations. The XPS data confirmed the successful substitution of $Ta^{5+}$ by $V^{5+}$ in the $\beta$-TaON lattice. The DFT results further revealed a reduction in effective electronic mass, enhanced dielectric permittivity, and favorable band edge alignments for overall water splitting. PEC performance improves at moderate vanadium levels due to efficient charge separation and reduced recombination rate. At higher loadings, phase segregation hinders interfacial transport, limiting the achievable photocurrent. Changes in open-circuit potential values with irradiance support the role of the local chemical environment in facilitating interfacial charge transfer. The study also revealed key limitations. At the vanadium concentrations of ≥15 at.%, secondary phases and deep-level defects decreased PEC performance, while structural disorder at high doping levels contributed to variability in the optical and electrochemical responses. These findings indicated that only moderate vanadium incorporation (≤10 at.%) could provide consistent PEC improvements. The results highlighted the potential of vanadium as an effective dopant for optimizing $\beta$-TaON under visible-light irradiation and defined the incorporation range required to achieve enhancements in the PEC response.

## 4 | Experimental Section

### 4.1 | Synthesis

Pristine and vanadium-doped $\beta$-TaON powders were synthesized according to the validated synthesis conditions reported elsewhere [49]. First, $Ta_2O_5$ (0.15 g, 99%, Alfa Aesar) was mixed with the appropriate amount of $V_2O_5$ (99.9%, Merck) according to the nominal atomic percentage in the Ta-V precursor mixture to obtain nominal vanadium doping concentrations of 0, 5, 10, 15, 20, and 25 at.%. Then, the powders were thoroughly ground using an agate mortar and a pestle and evenly spread in a corundum crucible. The well-homogenized mixture was subjected to ammonolysis in a horizontal tube furnace equipped with a $SiO_2$ tube under a localized gas delivery system [50]. The mixture was heated up to 900°C at a heating rate of 400 K•h$^{-1}$ and held at this temperature for 12 h under a flowing gas mixture of ammonia ($NH_3$, 10 L•h$^{-1}$, ≥99.95%, Air Liquide) and oxygen ($O_2$, 0.25 L·h$^{-1}$, 99.999%, Air Liquide). Following the high-temperature reaction, the sample was allowed to cool naturally to room temperature. The resulting samples were labeled as TVON0, TVON5, TVON10, TVON15, TVON20, and TVON25 according to their nominal vanadium doping concentrations.

### 4.2 | Characterization

X-ray diffraction (XRD) patterns were obtained using a PANalytical X'Pert Pro powder diffractometer operated with nickel-filtered Cu-$K_\alpha$ radiation at 40 kV and 30 mA. The powder diffraction data were collected in a Bragg–Brentano setup with a $\theta/\theta$-arrangement at ambient temperature over an angular range of $2\theta = 10$–80° with a step size of 0.020°. The microstructure of the samples was examined by scanning electron microscopy (SEM; Zeiss Gemini 982, Carl Zeiss) at an operating voltage of 5 kV. The nanostructure of the samples and their elemental distributions were characterized using a transmission electron microscope (TEM, JEM-2100F, JEOL) equipped with a CETCOR aberration corrector (CEOS GmbH), operating at an acceleration voltage of 80 kV and 200 kV, respectively. The surface chemical states of the samples were probed by X-ray photoelectron spectroscopy (XPS; JPS-9010MX, JEOL) using Mg-K$\alpha$ radiation. The ultraviolet-visible (UV–vis) diffuse reflectance spectra of the samples were measured using an Evolution 220 UV–vis spectrophotometer (Thermo Fisher Scientific).

### 4.3 | Photoelectrochemical (PEC) Performance Evaluation

PEC measurements were carried out to assess the light-induced charge transfer characteristics of photocatalyst powders immobilized onto conductive electrode surfaces. A DropSens µSTAT200 potentiostat was employed to perform the electrochemical tests. The working electrolyte volume was fixed to 100 µL, with an illuminated geometric area of 0.13 cm$^2$. Commercial screen-printed electrodes (SPEs) (DropSens 110) were used as the conductive substrates. Photocatalyst films were fabricated by dip-coating, following a modified procedure adapted from previous reports [51, 52]. Briefly, the photocatalyst powder was dispersed in a 1:1 ethanol-water mixture to a 1.0 mg•mL$^{-1}$ suspension, which was ultrasonicated for 30 min to ensure homogeneity. The resulting suspension was deposited onto the electrode surface, which was then covered with a glass beaker, and dried using a heat gun at 80 °C for 10 min. This coating-drying cycle was repeated three times to achieve a typical mass loading of approximately 0.18 mg•cm$^{-2}$. All PEC measurements were conducted in a 0.1 M $Na_2SO_4$ electrolyte solution, previously purged with nitrogen gas for 10 min. The working electrode was illuminated with simulated solar light (290–900 nm) at an intensity of $P_{solar}$ ~100 mW•cm$^{-2}$ (Solar Light Co.). A concentric carbon ring served as the counter electrode, while an Ag/AgCl electrode was used as the reference. All potentials were converted to the reversible hydrogen electrode (RHE) scale using the Nernst equation:

$$\text{E (vs. RHE)} = \text{E (vs. Ag/AgCl)} + 0.059 \times \text{pH} + 0.197 \quad (1)$$

It should be noted that, given the inherent limitations of dip-coated powder films on carbon SPEs, namely non-uniform illumination, possible representative series resistance, and uncertainty in particle-electrode contact, the PEC measurements in this work were interpreted strictly as screening-level trends within the material samples rather than as absolute performance indicators. All photoelectrochemical measurements, including linear sweep voltammetry (LSV), applied-bias photon-to-current efficiency (ABPE), and chronoamperometry (CA), were performed under an illumination intensity of 100 mW cm$^{-2}$. LSV was conducted at a scan rate of 10 mV•s$^{-1}$. ABPE was used as an internal metric to compare samples under potentiostatic three-electrode conditions and calculated as:

$$\text{APBE} = \{[\text{Jphoto} \times (1.23 - \text{Ebias})] / \text{Psolar}\} \times 100\% \quad (2)$$

where $J_{photo}$ was the steady-state photocurrent density, $E_{bias}$ was the applied electrode potential expressed *vs.* RHE (V), and $P_{solar}$ is the irradiance [53]. CA measurements were carried out at 1 V *vs.* RHE under both dark and light conditions, first in the supporting electrolyte alone and then after the addition of 5 mM potassium ferrocyanide. In addition, the open-circuit potential (OCP using a PalmSens4 potentiostat) of the coated electrodes was monitored as a function of incident light intensity (30, 50, 80, 100, and 150 mW•cm$^{-2}$). The photocurrent onset potential was defined as the applied potential at which the net photocurrent density (illuminated minus dark current) obtained from LSV reaches 0.03 mA•cm$^{-2}$ and subsequently increases monotonically with increasing bias. This threshold-based criterion minimizes ambiguity arising from capacitive currents and contact-related variability, which are intrinsic to dip-coated powder electrodes on carbon SPEs, and enables reliable relative comparison across the V-doping series.

### 4.4 | Density Functional Theory (DFT) Calculations

Density functional theory (DFT) calculations were performed using the Vienna Ab Initio Simulation Package (VASP) with projector augmented-wave (PAW) pseudopotentials. Structural relaxations were carried out using the SCAN meta-GGA functional, while electronic and optical properties were evaluated with the Tran–Blaha modified Becke–Johnson (TB-mBJ) potential. All calculations were spin-polarized, and vanadium doping was modeled using Ta-site substitution in supercells with compositions corresponding to the experimental range. Full computational parameters, including k-point sampling, energy cutoffs, supercell configurations, and phonon calculation details, are provided in the Supporting Information.

**Acknowledgements**

This work was funded by a Novo Nordisk Foundation RECRUIT Grant (NNF23OC0079059). This work was also supported in part by the World Research Hub (WRH) Program of the Institute of Science Tokyo and the "Invitational Fellowship for Research in Japan" Program of the Japan Society for the Promotion of Science (S22063). This work was also supported in part by the Meiji University International Collaborative Research Promotion Project (MU-OSRI-ICRPP2023-206). MH would like to thank Prof. Martin Lerch for providing access to his research laboratory at Technische Universität Berlin, Germany. The authors would like to thank Dipl. Phys. Christoph Fahrenson from ZELMI, TU Berlin, Germany, and Reiko Shiozawa from Shinshu University, Japan, for their technical support in SEM and XPS analyses, respectively. KY would like to thank the Japan Society for the Promotion of Science (KAKENHI Grant Number. JP23K04373) and the Institute for Materials Research, Tohoku University, Japan (GIMRT Program, proposal no. 202311-RDKGE-0001).

**Funding**

This work was funded by a Novo Nordisk Foundation RECRUIT Grant (NNF23OC0079059). This work was also supported in part by the World Research Hub (WRH) Program of the Institute of Science Tokyo and the "Invitational Fellowship for Research in Japan" Program of the Japan Society for the Promotion of Science (S22063). This work was also supported in part by the International Collaborative Research Promotion Project of Meiji University (MU-OSRI-ICRPP2023-206). KY would like to thank KAKENHI Grant of Japan Society for the Promotion of Science (JSPS) Grant no. JP23K04373 and the Institute for Materials Research, Tohoku University, Japan (GIMRT Program, proposal no. 202311-RDKGE-0001).

**Conflicts of Interest**

The authors declare no conflicts of interest.

**Data Availability Statement**

The data that support the findings of this study are available within the article and its Supporting Information, or from the corresponding author upon reasonable request.

## Supporting Information

Additional supporting information can be found online in the Supporting Information section.

**Supporting File**: smll72835-sup-0001-SuppMat.docx.